\input{aipcheck}

\documentclass[
    ,final            
  ]
  {aipproc}

\layoutstyle{6x9}

\newcommand	     \cc              {cm$^{-3}$}
\newcommand	     \mic              {$\mu$m}

\begin{document}

\title{Dust-Gas Interaction in SNR 1987A}


\author{Eli Dwek}{
  address={Observational Cosmology Lab, NASA Goddard Space Flight Center, Greenbelt, MD 20771\\ e-mail: eli.dwek@nasa.gov}
}

\author{Richard G. Arendt}{
  address={CRESST/UMBC, NASA Goddard Space Flight Center, Greenbelt, MD 20771}
}
%

\begin{abstract}

Multiwavelength observations of SNR~1987A show that its morphology is rapidly changing at X-ray, radio, and optical wavelengths as the blast wave from the explosion expands into the circumstellar equatorial ring.    
Infrared emission arises from the interaction of dust grains with the hot X-ray emitting gas. We show that the IR emission provides important complementary information on the interaction of the SN blast wave with the circumstellar equatorial ring that cannot be obtained at any other wavelength. 
\end{abstract}

\maketitle


\section{Introduction: The SN to SNR Evolution of SN1987A}
On February 23, 1987 a supernova, designated SN~1987A, exploded in the LMC. It has since evolved from a supernova
(SN) dominated by the emission from the radioactive decay of $^{56}$Co ($\tau_{1/2}= 77.3$~d), $^{57}$Co ($\tau_{1/2}= 271.8$~d), and $^{44}$Ti ($\tau_{1/2}= 63$~yr) in
the ejecta to a supernova remnant (SNR) whose emission is dominated by the interaction of its blast wave with the immediate surrounding medium. The circumstellar medium exhibits a complex structure, with the closest and densest feature being an inner equatorial ring (ER) (e.g. Burrows et al. 1995; McCray 2007, Figure 1). Ongoing observations of the SN with
the {\it Hubble Space Telescope} ({\it HST}) have revealed spectacular images of ``hot spots'' strung
along the ER like beads on a necklace, with new ``hot spots'' continuing to appear as the
blast wave propagates into the ring (Pun et al. 1997). These ``hot spots'' appear to be fingerlike
protrusions, the first ER material to be hit by the blast wave (McCray 2007; Figure 2). Images of the ER obtained
at 0.3-8 keV with the {\it Chandra X-ray Observatory} (Park et al. 2005b), at 12 mm with the Australian Telescope
Compact Array (ATCA) at the Australian Telescope Facility, and at 11.7 and 18.3 $\mu$m with
the T-ReCS mid-infrared (IR) imager at the Gemini-S 8m telescope (Bouchet et al. 2004) show
an overall similar morphology, characterized by similar mean radii and surface brightness
distribution. A multiwavelength comparison of the morphological changes of the ER is presented in McCray (2007; Figure 6). 

The X-ray emission is thermal emission from the very hot plasma, shock heated to temperatures in excess of $\sim10^7$ K by advancing and reflected shocks generated by the SN blast
wave (Park et al. 2007; this volume). The optical emission arises from the gas that is shocked by the blast wave transmitted
through the dense knots in the ER (Kirshner et al. 2007; this volume; Pun et al. 2007; this volume), and the radio emission is synchrotron radiation from
shock-accelerated electrons. The mid-IR emission comprises line and continuum emission.
The lines originate from the optically bright dense knots, whereas the continuum, which
dominates the spectrum, is thermal emission from collisionally-heated dust that is immersed in the X-ray emitting plasma. This dust was formed in the quiescent outflow of the progenitor star before it exploded.

The interaction of the SN blast wave with  the ER is causing rapid changes in the morphology and the radiative outputs from the SNR across all wavelength regions on a dynamical time scale of about a year. Here we report on the recent IR evolution of SNR1987A, and what we can learn from the comparison of the IR and X-ray fluxes  on the physical condition of the X-ray emitting plasma and gas-grain interaction in the shocked gas. 
 
\section{The Infrared Diagnostic of Dusty X-ray Plasma}

The morphological similarity between the X-ray and mid-IR images of SN~1987A suggest that  the IR emission arises from dust that is collisionally heated by the X-ray emitting gas. Simple arguments, presented below, show that under certain conditions the IR luminosity and spectrum of a dusty plasma can be used as a diagnostic for the physical conditions of the gas and details of the gas-grain interactions. Details of the arguments can be found in Dwek (1987) and Dwek \& Arendt (1992).
 
\subsection{The Dust Temperature as a Diagnostic of Electron Density}
We will assume that the dust is primarily heated by electronic collisions with the ambient gas.    
The collisional heating rate, ${\cal H}$ of a dust grain of radius $a$ embedded in a hot plasma with electron density, $n_e$ and gas temperature, $T_g$ is given by:
\begin{equation}
\label{hcoll}
{\cal H} = \pi a^2\ n_e\ v_{th}\ E_{dep}
\end{equation} 
where $v_{th} \propto T_g^{1/2}$ is the thermal velocity of the electrons, and $E_{dep}$ is the average energy deposited by electronic collisions in the dust. If most electrons are stopped in the dust then $E_{dep}$ is, on average, equal to the thermal energy of the electrons, that is, $E_{dep} \propto T_g$. On the other hand, if most incident electrons penetrate the grains, then 
\begin{equation}
\label{ }
E_{dep} \approx \left( {dE\over dx}\right)\, a\ \ \propto \ \ a\, E^{-1/2} \ \  \propto \ \ a\, T_g^{-1/2}
\end{equation}
where the electron stopping power in the solid is defined as $\rho^{-1}(dE/dx)$, where $\rho$ is the mass density of a grain. At the energies of interest here the electronic stopping power has an energy dependence of $\sim E^{-1/2}$  (Iskef et al. 1983). The functional dependence of the grain heating rate on gas density and temperature is then given by:

\begin{eqnarray}
{\cal H}  & \propto &   a^2\ n_e\ T_g^{3/2}  \qquad \ \ \ {\rm when\ electrons\ are\ stopped\ in\ the\ grain} \\ \nonumber
 & \propto &  a^3\ n_e \qquad \qquad \ \ \ {\rm when\ electrons\ penetrate\ the\ grain}
\end{eqnarray}

The radiative cooling rate, ${\cal L}$, of the dust grain with temperature $T_d$ by IR emission is given by:
\begin{eqnarray}
\label{ }
{\cal L} & = & \pi a^2\ \sigma T_d^4\ \left< Q \right> \\ \nonumber
& \propto & \pi a^3\ \sigma T_d^{4+\beta}
\end{eqnarray}
where $\sigma$ is the Stefan-Boltzmann constant, and $\left< Q \right> \propto a\, T_d^{\beta}$ is the Planck-averaged value of the dust emissivity, $Q(\lambda) \propto \lambda^{-\beta}$, where the value of emissivity index, $\beta$, is $\approx 1-2$.

In equilibrium, ${\cal H} = {\cal L}$, and the dust temperature dependence on plasma density and temperature can be written as:
\begin{eqnarray}
T_d & \propto & \left({n_e \over a}\right)^{\gamma}\, T_g^{3\gamma/2} \qquad \ \ \ {\rm when\ electrons\ are\ stopped\ in\ the\ grain} \\ \nonumber
 & & \\ \nonumber
 & \propto & n_e^{\gamma} \qquad  \qquad \ \ \ \qquad {\rm when\ electrons\ penetrate\ the\ grain} 
\end{eqnarray}
where $\gamma \equiv 1/(4+\beta)$. These simple arguments show that when the gas temperature is sufficiently high, and the grain size sufficiently small so that most electrons penetrate the grain, the dust temperature depends only on plasma density. 

\begin{figure}
      \vspace{-5.0in} 
  \includegraphics[width=5.0in]{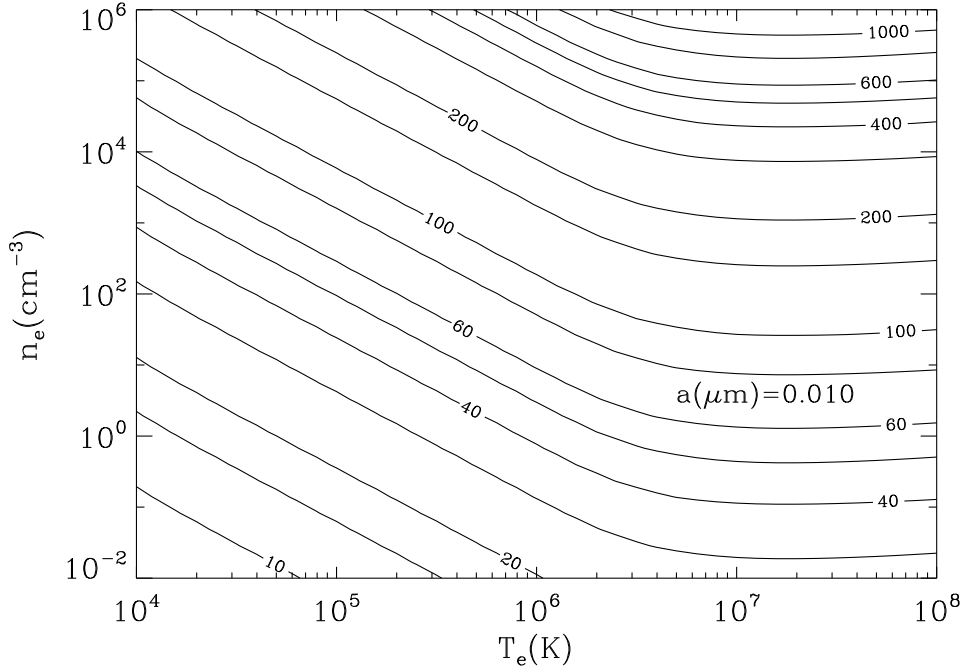} 
  \caption{{\footnotesize Contour plot of the equilibrium temperature of 0.01~\mic\ silicate grains as a function of electron density and temperature. Above temperatures of $\sim 5\times10^6$~K the grains become transparent to the incident electrons, and the dust temperature is only a function of electron density. We note here that a similar figure (Figure 15 in Bouchet et al. 2006) was mislabeled, and actually corresponds to contour levels of silicate dust temperature for grain radius of $a=0.003$~\mic.}}
  \label{tempeq}
\end{figure}

Figure \ref{tempeq} depicts contour levels of the dust temperature as a function of electron density and temperature for 0.01~\mic\ silicate grains. The figure shows that above temperatures of $\sim 5\times 10^6$~K most of the dust particles penetrate the grains and the dust temperature is essentially determined by the electron density. Under these conditions, the IR spectrum from the collisionally-heated grains becomes an excellent diagnostic of the density of the  X-ray emitting gas.

\begin{figure}
\begin{tabular}{cc}
  \includegraphics[width=3.0in]{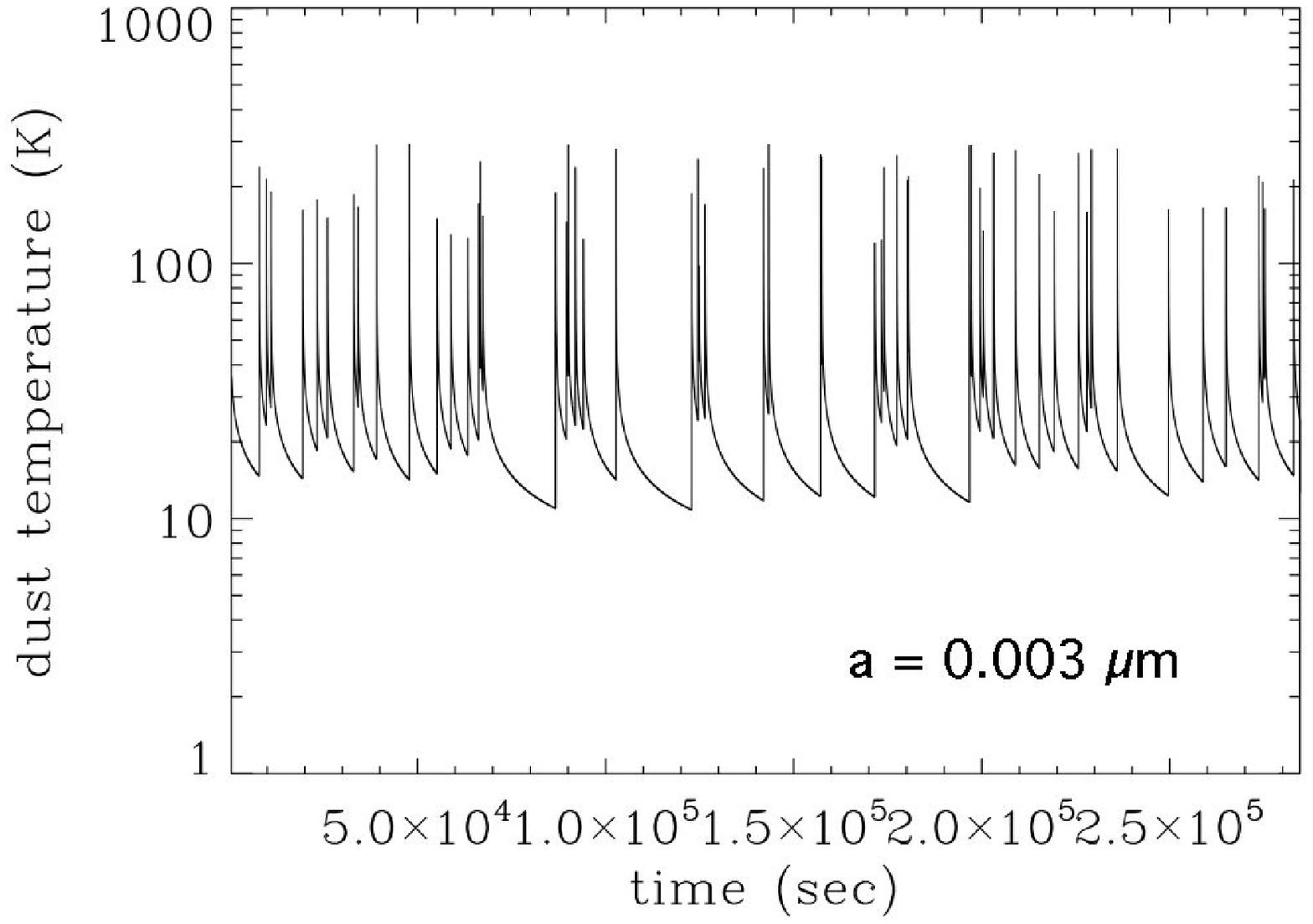} 
  \includegraphics[width=3.0in]{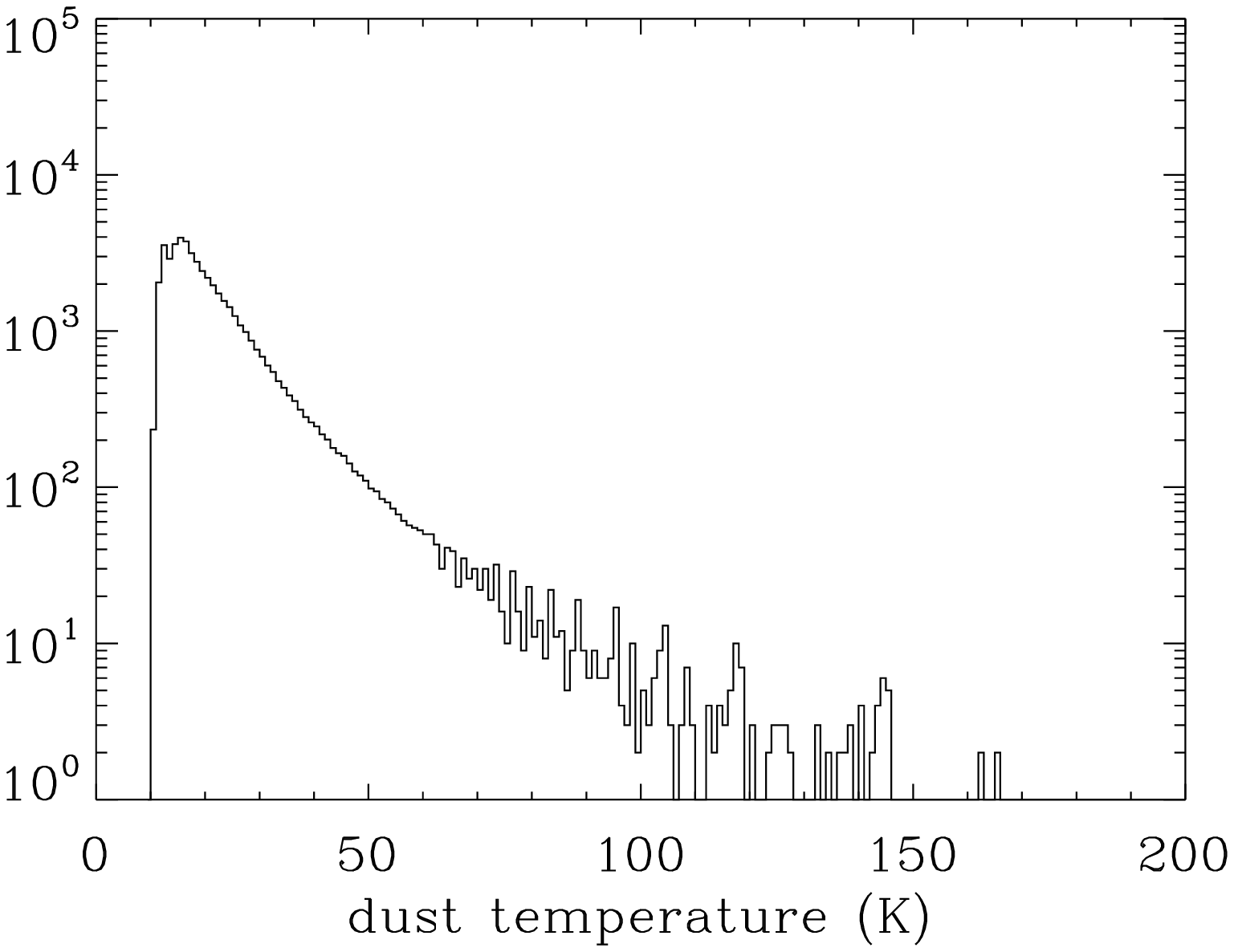}\\
  \includegraphics[width=3.0in]{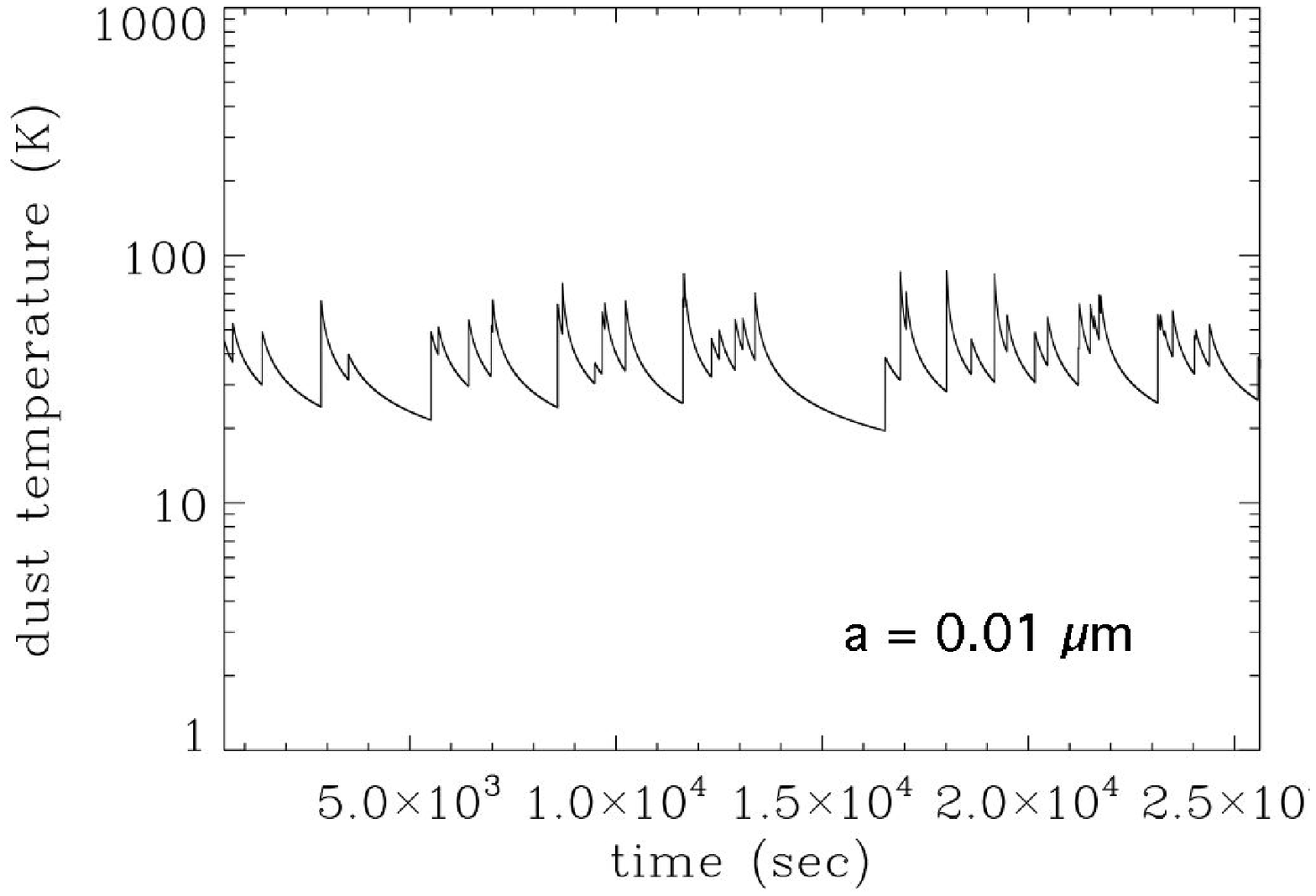} 
  \includegraphics[width=3.0in]{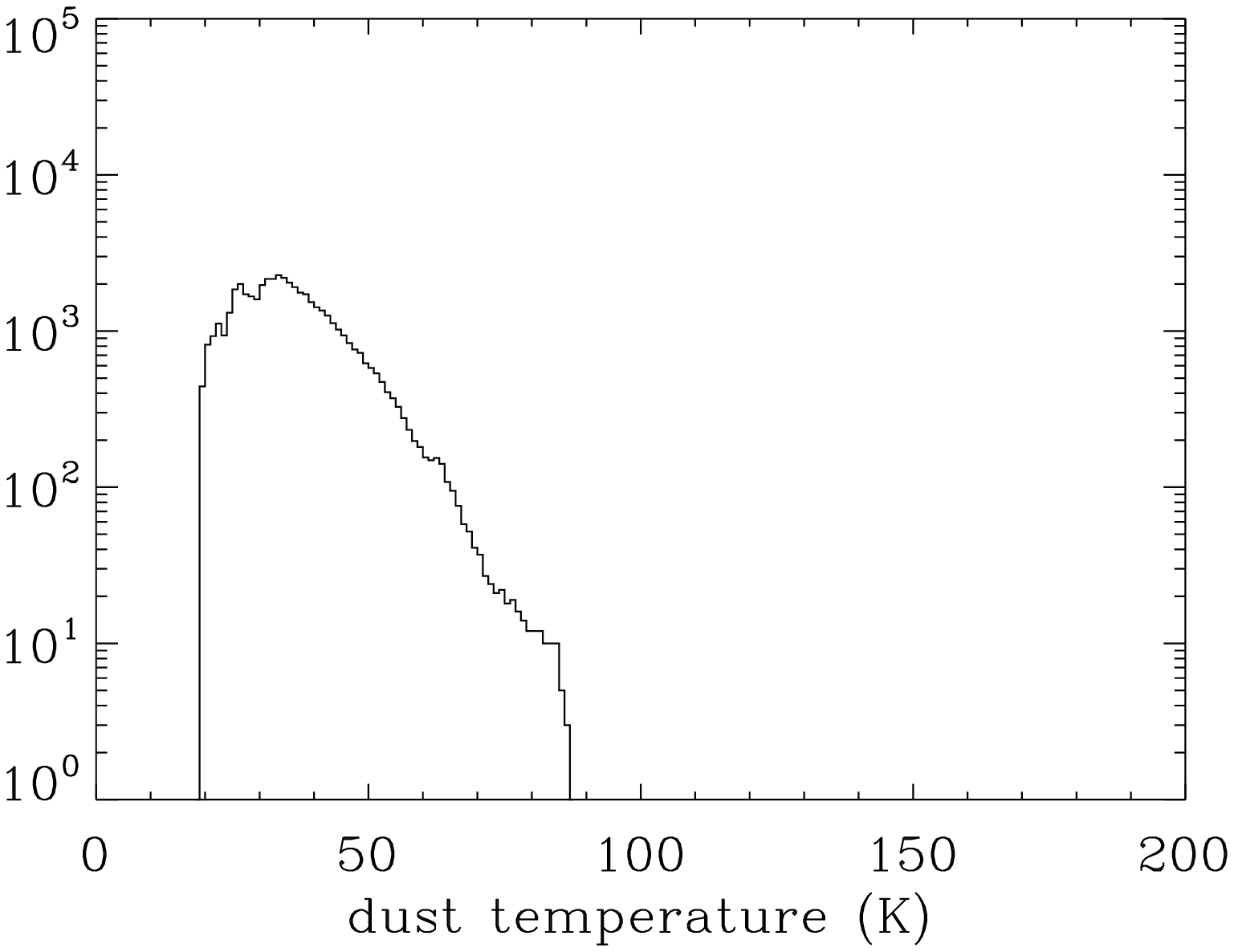}\\
    \includegraphics[width=3.0in]{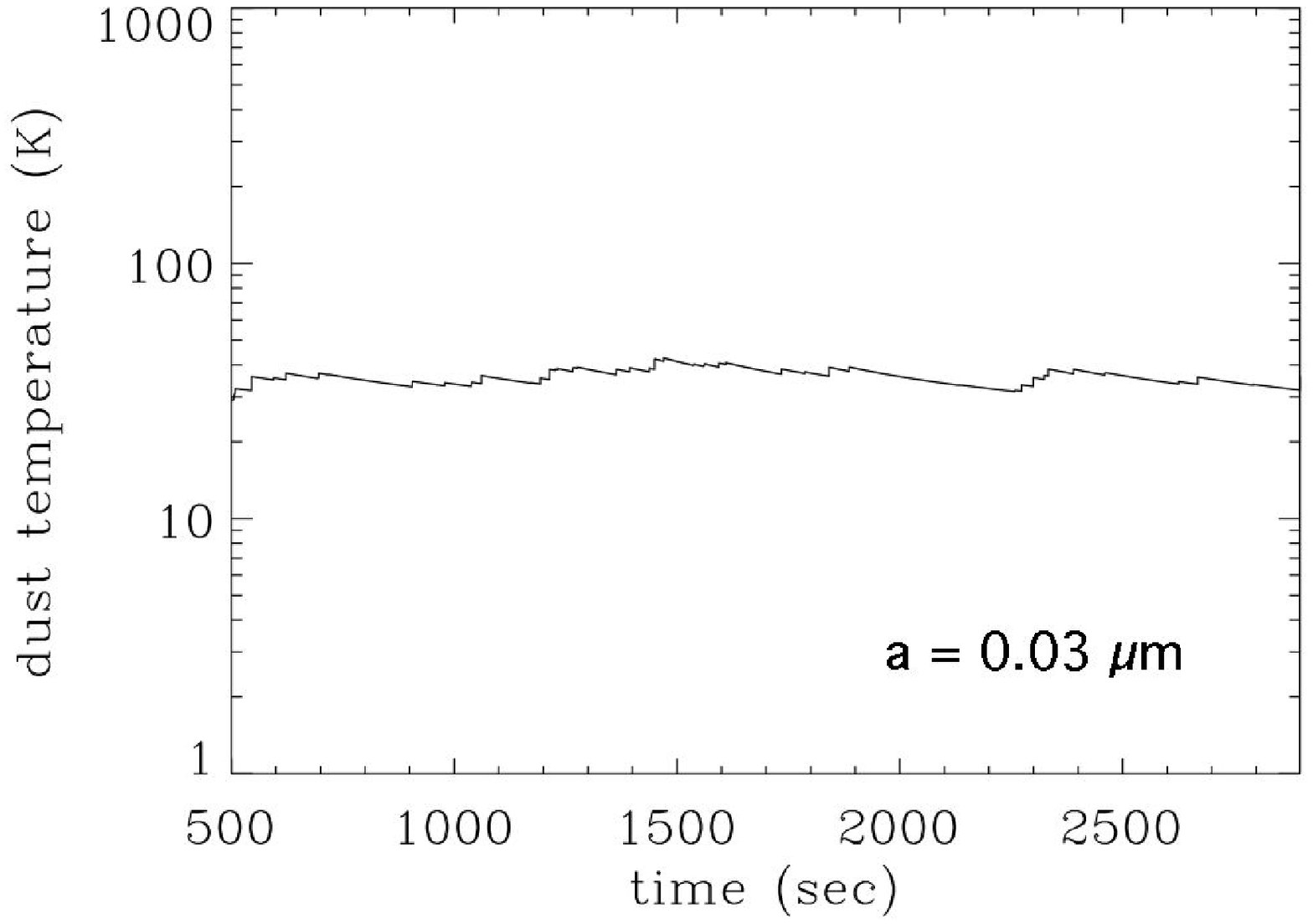} 
   \includegraphics[width=3.0in]{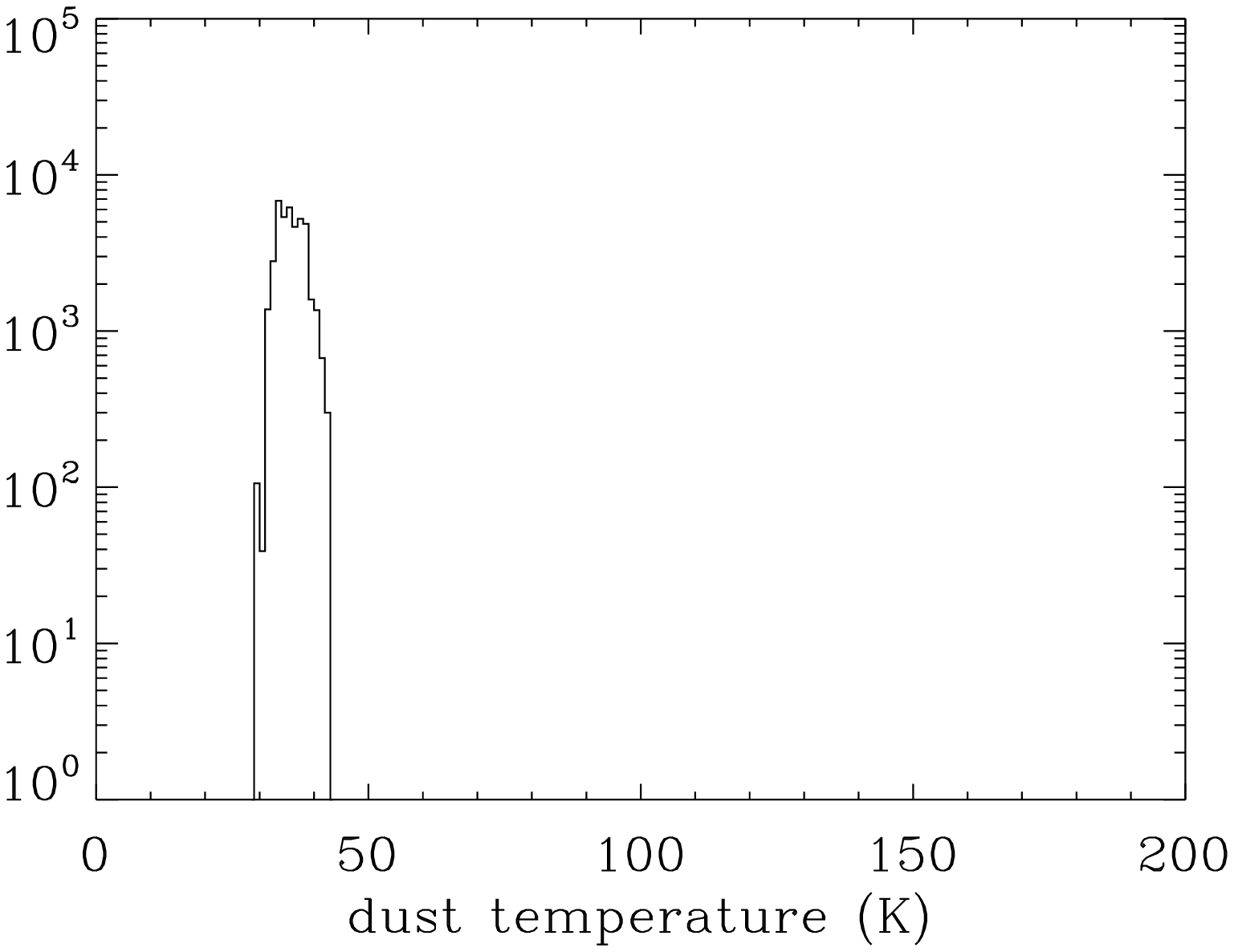}
  \caption{{\footnotesize The stochastic heating of silicate grains in a hot X-ray emitting gas characterized by a temperature of $T_g=10^6$~K, and electron density $n_e=1$~\cc\ for dust grains of different radii. {\bf Left column:} The temperature fluctuations as a function of time. {\bf Right column:} The histogram of the fluctuations. As the grain size increases, the fluctuations get smaller, and the probability distribution of dust temperatures becomes strongly peaked around the equilibrium temperature of $\sim 40$~K.}}
\end{tabular}
\label{tfluc}
\end{figure}

\subsection{The Stochastic Heating of Grains by Electronic Collisions}
When dust grains are sufficiently small, a single electronic collision can deposit an amount of energy in the dust that is significantly larger that its enthalpy, causing a surge in dust temperature. If additionally, the time interval between successive electronic collisions is larger that the dust cooling time, the grain temperature will be fluctuating with time.  Figure \ref{tfluc} depicts a simulation of the stochastic heating of 0.003, 0.01, and 0.03~\mic\ silicate grains immersed in a hot X-ray emitting gas characterized a temperature $T_g = 10^6$~K, and an electron density $n_e=1$~\cc. The left column shows the temperature fluctuations as a function of time, and the right column the histogram of the grain temperature. As the grain size increases, the fluctuations get smaller, and the histogram becomes strongly peaked around the equilibrium dust temperature, $\sim 38$~K in  this example.

\begin{figure}[b]
\begin{tabular}{cc}
  \includegraphics[width=2.5in]{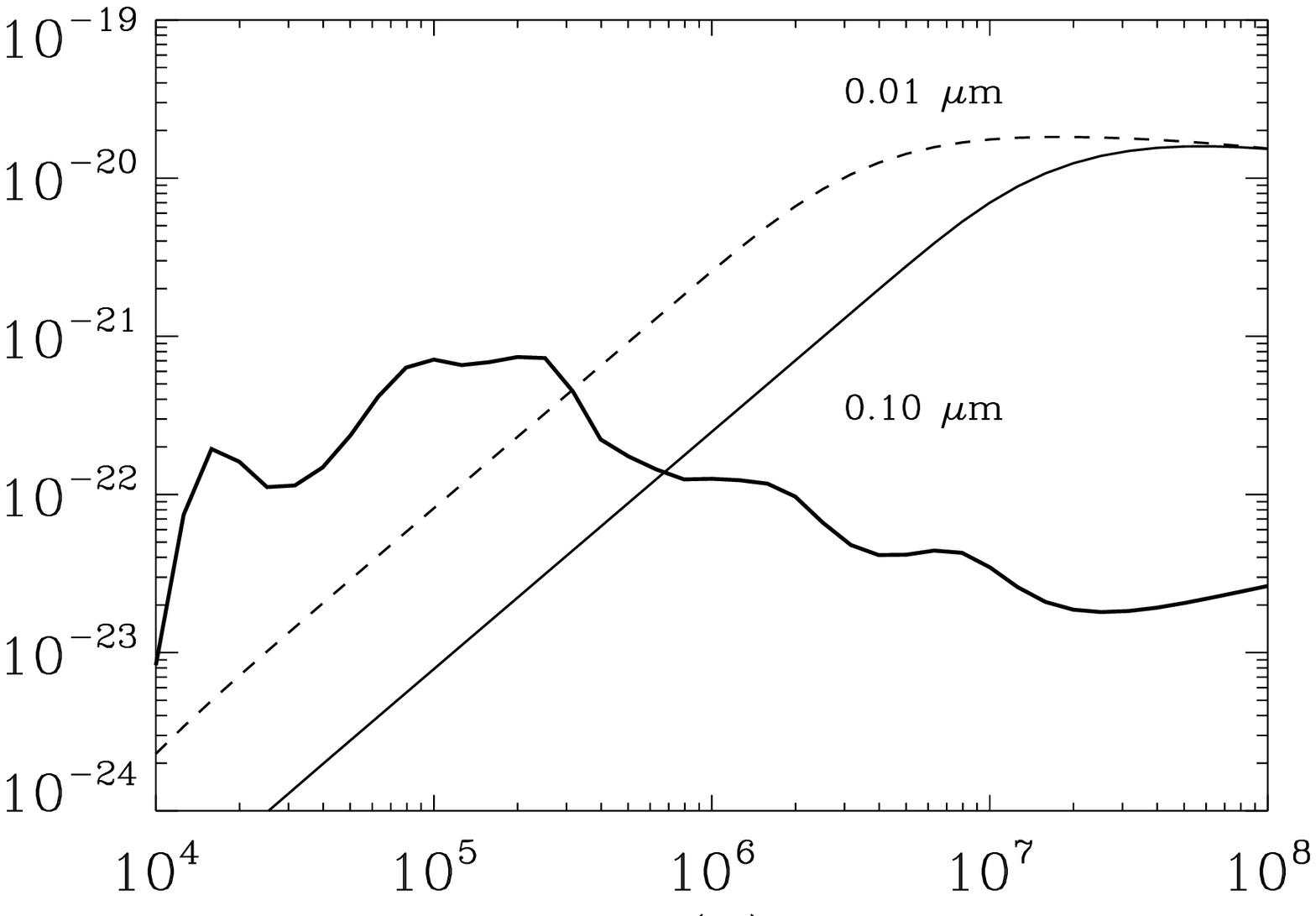}
  \hspace{0.3in} 
   \includegraphics[width=2.5in]{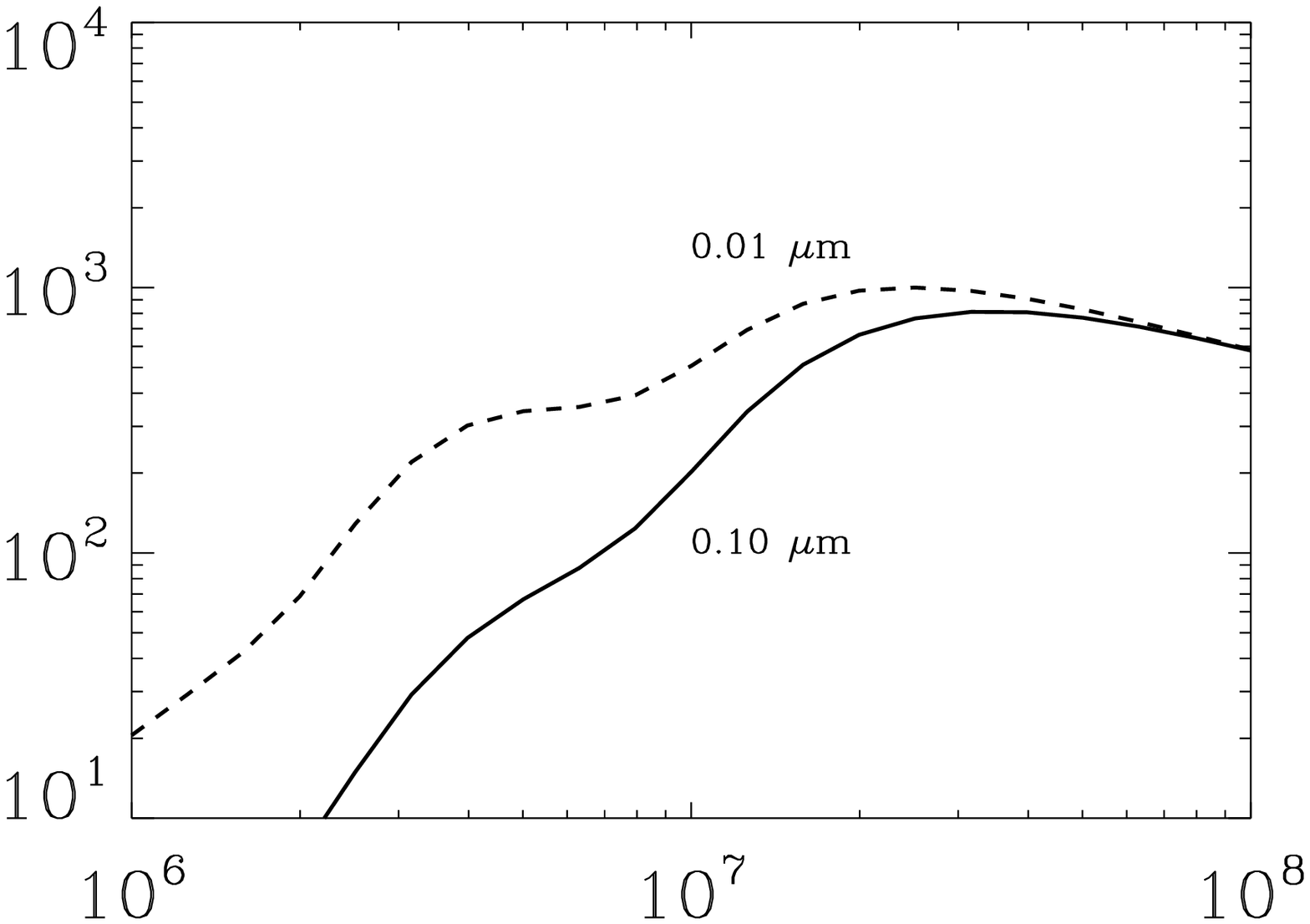} 
   \vspace{0.2in} 
  \caption{{\footnotesize {\bf Left panel}: The cooling function of a dusty plasma via atomic processes (thick solid line) and via gas grain collisions calculated for a dust-to-gas mass ratio of 0.01, assuming a single population of 0.01~\mic\ dust grains (dashed line) and a single population of 0.10~\mic\ grains (solid line). {\bf Right panel}: The $IRX$ flux ratio, given by the ratio of the dust and gas cooling functions for the two dust populations in the left panel. }}
  \end{tabular}
  \label{irx}
\end{figure}

\subsection{The Infrared to X-ray Flux Ratio}
Another important diagnostic of a dusty plasma is its $IRX$ flux ratio, defined as the ratio of the IR to X-ray fluxes emitted by the gas (Dwek et al. 1987). If the dust is collisionally-heated by the gas then the total IR flux, $F_{IR}$, emitted from a gas volume $V$ is proportional to $n_e\ n_d\ \Lambda_d(T_g)\ V$, where $n_d$ is the number density of dust particles, and $\Lambda_d(T_g)$ is a cooling function of gas via gas--grain collisions. The total X-ray flux, $F_X$, from the same volume is proportional to $n_e^2\ \Lambda_g(T_g)\ V$, where $\Lambda_g(T_g)$ is cooling function of the gas via atomic processes. 
\begin{equation}
\label{ }
IRX \equiv \left({n_d\over n_e}\right) \, {\Lambda_d(T_g)\over \Lambda_g(T_g)}
\end{equation}

For a given dust-to-gas mass ratio, that is a fixed $(n_d/n_e)$ ratio, the $IRX$ flux ratio depends only on plasma temperature. Figure \ref{irx} (left panel) shows the behavior of the atomic cooling function of a gas of solar composition as a function of gas temperature. Also shown in the figure is the gas cooling function via gas--grain collisions for a gas with a dust-to-gas mass ratio of 0.01 and a single-sized dust population with radii of 0.01~\mic\ (dashed line) and 0.1~\mic\ (solid line). The right panel of the figure presents the $IRX$ flux ratio for the same conditions. The figure shows that this ratio varies between $\sim 10^2$ and $10^3$ in the $ \sim 5\times10^6$ to $1\times 10^8$~K temperature region. Any deviation from this ratio will suggest that the the dust-to-gas mass ratio is either depleted of overabundant with respect to the reference value adopted in the calculations.

\section{{\it Spitzer} Infrared Observations of SNR~1987A}
\subsection{ The IR spectrum: constraining the plasma density}
\begin{figure}[t]
\begin{tabular}{cc}
  \includegraphics[width=3.0in]{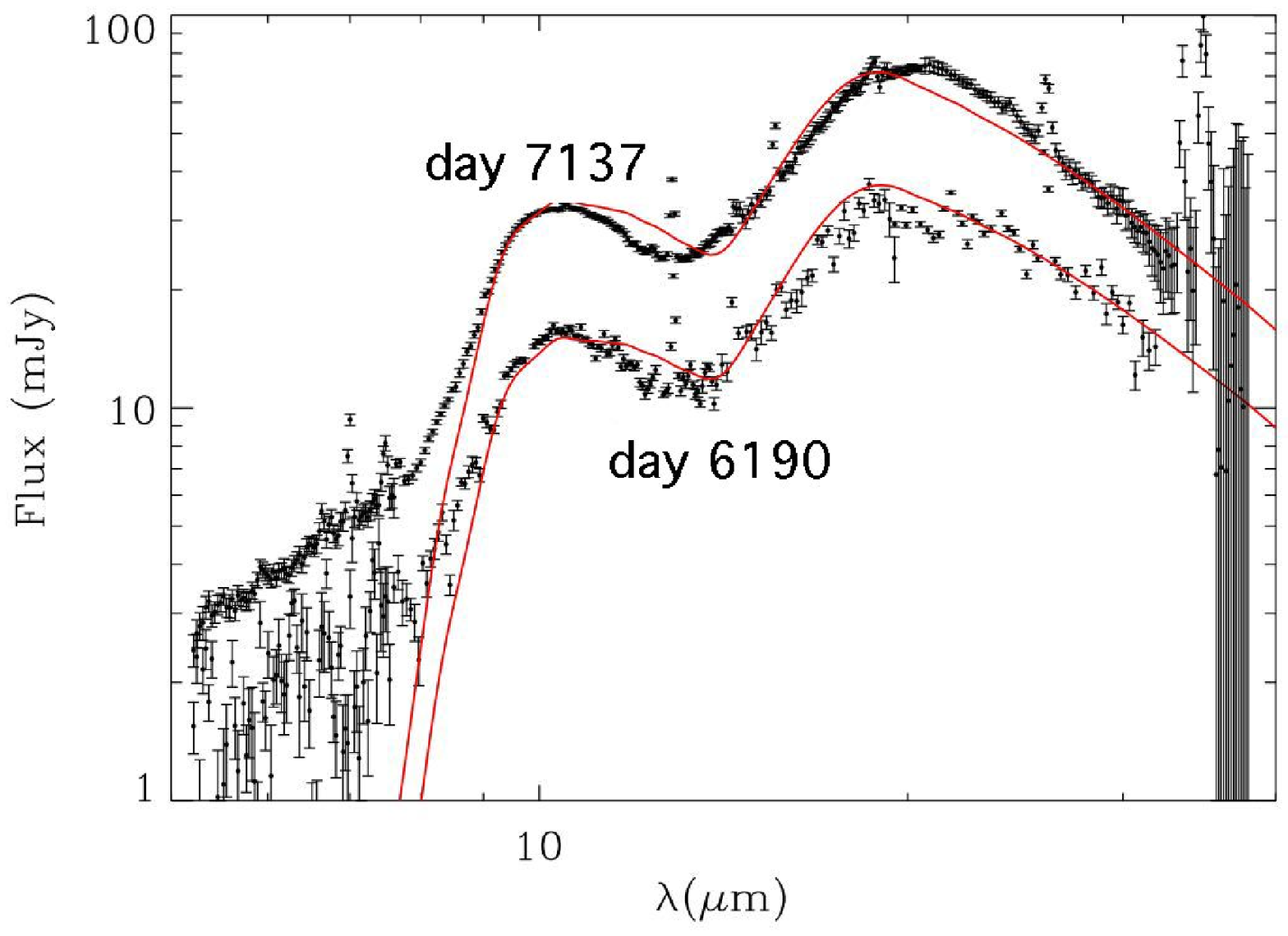}  
   \includegraphics[width=3.0in]{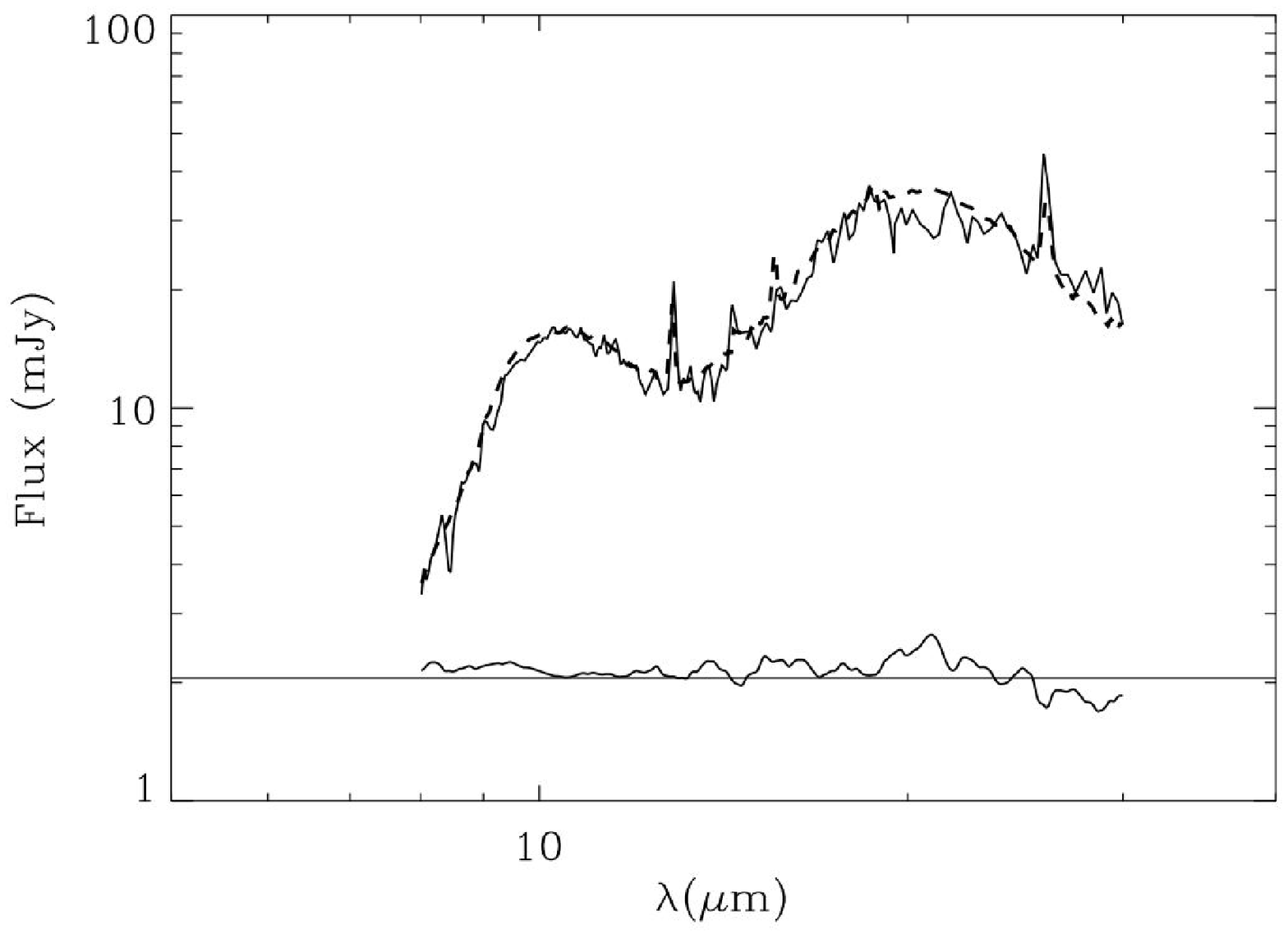}
   \vspace{-1.7in} 
  \caption{{\footnotesize {\bf Left panel}: The evolution of the IR spectrum of SN1987A from April 2, 2004 (day 6190 since the explosion) to September 8, 2006 (day 7137) taken with the {\it Spitzer} IRS (Bouchet et al. 2007; Arendt et al. 2007). {\bf Right panel}: The smoothed spectra for days 6190 (solid line) and day 7137 (dashed line), normalized to the same brightness. The lower curve shows the ratio between the to spectra, with the horizontal line being the mean value. The figure shows that the dust spectrum increased by a factor of two between the two epoch, retaining essentially an identical spectrum corresponding silicate grains radiating at an equilibrium temperature of $\sim 180$~K. }}
  \end{tabular}
  \label{specvol}
\end{figure}

Figure \ref{specvol} shows the $\sim 5 - 30$~\mic\ low resolution spectra of SNR~1987A taken on days 6190 and 7137 after the explosion with the Infrared Spectrograph (IRS) on board the {\it Spitzer} satellite. Analysis of the spectrum taken on day 6190 revealed that the IR emission originated from silicate grains radiating at a temperature of $\sim 180^{+20}_{-15}$~K (Bouchet et al. 2006). These circumstellar grains were formed in the quiescent outflow of the progenitor star before it exploded. 
X-ray observations of the ER suggest plasma temperatures of $\sim 2\times 10^7$~K (Park et al., this volume). At this temperature, dust particles with radii below $\sim 0.1$~\mic\ are transparent to the incident electrons, and the dust temperature can be used to measure the plasma density.  Figure \ref{tempeq} shows that dust temperatures of $\sim 180$~K can
be achieved for electron densities of $\sim 800$~\cc\ when $T_e > 5 \times 10^6$~K. This density is in excellent agreement
with the range of gas temperatures inferred from the {\it Chandra} X-ray observations (Park et al. this volume). 

\subsection{The IRX flux ratio: constraining dust abundance} 
An upper limit on the $IRX$ flux ratio can be obtained by comparing the observed IR flux from SNR~1987A to the observed X-ray flux in the 0.20-4.0~keV band. The total IR flux on day 6190 was $7.7\times 10^{-12}$~erg~cm$^{-2}$~s$^{-1}$ (Bouchet et al. 2007), comparable to the 0.20-4.0~keV X-ray flux of $7.2\times 10^{-12}$~erg~cm$^{-2}$~s$^{-1}$, giving an $IRX$ value $< 1$. 

The low value of the $IRX$ flux ratio, compared to the expected one of $10^{2-3}$, suggests that the dust 
is severly depleted compared to standard dust-to-gas mass ratio in the
LMC, either because the dust condensation efficiency in the wind of the pre-SN star
was very low, or that the dust is being destroyed in the hot X-ray gas.

\subsection{The evolution of the IR Spectrum of SNR~1987A: Evidence for grain destruction}
Recent {\it Spitzer} observations taken on  September 8, 2006 (day 7137 since the explosion) shows that the IR flux has increased by a factor of 2 since day 6190 (see Figure \ref{specvol}). The figure shows that the dust composition and temperature remained unchanged during the two observing periods. These observations provide strong constraints on the interaction of the SN blast wave with the equatorial ring, suggesting that the density of the X-ray emitting gas remained essentially unchanged at $\sim 800$~\cc\ between these two epochs.

In comparison, the X-rays from the ER have increased by a factor of 3, during the same period (Park et al. this volume). 
If grains were not destroyed, we would expect the IR intensity to increase by a similar factor. {\it The smaller increase in the IR, e.g. the decline in the $IRX$ ratio, is a strong indicator that we are for the first time witnessing the actual destruction of dust in a shock on a dynamical timescale!}
 Grain destruction becomes important on time scales of $\sim$1 yr, if the
initial radii of the silicate dust grains are smaller than $\sim$ 50 \AA\ (Bouchet et al. 2006).

\section{Summary}
The rapid changes that have been observed in the optical ({\it Hubble}), X-ray ({\it Chandra}), and mid-IR (Gemini South) morphology are all manifestations of
the interaction of the SN blast wave with the complex structure of the ER.   
The infrared observations provide important complementary information on the evolution of the interaction of the SN blast wave with the ER that cannot be obtained at any other wavelength. 
In particular, the mid-IR images have established that the IR emission originates from the interaction of the SN blast wave with the circumstellar material. {\it Spitzer} spectral observations on day 6190 after the explosion revealed that the radiating dust consists of silicate dust grains radiating at a temperature of $\sim 180$~K.  Subsequent observations on  day 7137 revealed that the IR flux increased by a factor of $\sim 2$, maintaining the same dust composition and temperature. These observations constrain the plasma density to be about $\sim 800$~\cc, and to remain roughly constant during the two epochs of IR observations. Comparison with the X-ray observations over the similar period show that the dust is severely depleted in the X-ray emitting gas, compared to standard LMC abundances, and that the dust is being destroyed on an evolutionary time scale ($\sim 1$~yr) in the shocked gas.

\begin{theacknowledgments}
E.D. thanks Dick McCray, Stefan Immler, and Kurt W. Weiler for organizing this stimulating conference in the wonderful surrounding of Aspen, CO, in celebration of the 20th birthday of SN~1987A. 
This work is based in part on observations made with the
{\it Spitzer} Space Telescope, which is operated by the Jet Propulsion Laboratory, California Institute of Technology, under a contract with NASA.
E.D. acknowledges partial support from HST grant GO-9114
for the Supernova INtensive
Survey (SInS: Robert Kirshner, PI), and by NASA OSS LTSA-2003-0065. 
The work of R.G.A. was supported by a grant awarded to Spitzer Cycle 3 proposal ID 30067.
\end{theacknowledgments}

\bibliographystyle{aipprocl} 


\begin{thebibliography}{9}


\bibitem[Bouchet et al. (2004)]{Bou04} Bouchet, P., et al., 2004, {\it ApJ}, 611, 394 

\bibitem[Bouchet et al. (2006)]{Bou06} Bouchet, P., et al., 2006, {\it ApJ}, 650, 212 
\bibitem[Burrows et al. (1995)]{Bur95} Burrows, C.J., et al., 1995, {\it ApJ}, 452, 680

\bibitem[Dwek (1986)]{Dwe86} Dwek, E., 1986, {\it ApJ}, 302, 363 

\bibitem[Dwek (1987)]{Dwe87a} Dwek, E., 1987, {\it ApJ}, 322, 812

\bibitem[Dwek et al. (1987b)]{Dwe87b} Dwek, E., et al., 1987, {\it ApJ}, 320, L27 

\bibitem[DA (1992)]{DA92} Dwek, E., \& Arendt, R. G. 1992, {\it Ann. Rev. Astron. \& Astroph.}, 30, 11


\bibitem[Iskef, Cunningham, \& Watt (1983)]{Isk83} Iskef, H., Cunningham, J. W., \& Watt, D. E. 1983, Phys. Med. Biol., 28, 535

\bibitem[K (2007) ]{K2007} Kirschner, R., et al. 2007, this volume
\bibitem[M (2007)]{McC2007} McCray, R., et al. 2007, this volume


\bibitem[P (2007)]{Park} Park, S., et al. 2007, this volume
\bibitem[Park et al. (2005b)]{Par05b} Park, S., et al., 2005, {\it ApJ}, 634, L73  


\bibitem[Pun et al. (2002)]{Pun02} Pun, C.S.J., et al., 2002, {\it ApJ}, 572, 906 


\end{thebibliography}




\end{document}